\documentclass[useAMS,usenatbib]{mn2e}

\usepackage{amsmath}
\usepackage{graphicx}
\usepackage{natbib}
\bibpunct{(}{)}{;}{a}{,}{,}

\newcommand{\msun}{\mbox{M$_\odot$}}

\title[Mechanical Properties]{Mechanical Properties of non-accreting Neutron Star Crusts}
\author[K. Hoffman and J. Heyl]{Kelsey Hoffman$^{1,2}$\thanks{email: kelsey@cita.utoronto.ca} and Jeremy Heyl$^{2}$\thanks{Email: heyl@phas.ubc.ca; Canada Research Chair}\\
$^{1}$Canadian Institute for Theoretical Astrophysics, University of Toronto, 60 St. George Street, Toronto, Ontario, M5S 3H8, Canada \\
$^{2}$Department of Physics and Astronomy, University of British
Columbia, 6224 Agricultural Road,\\ 
~~Vancouver, British Columbia, V6T 1Z1, Canada}

\begin{document}

\date{Accepted ---. Received ---; in original form ---}

\pagerange{\pageref{firstpage}--\pageref{lastpage}} \pubyear{2012}

\maketitle

\label{firstpage}

\begin{abstract}
  The mechanical properties of a neutron star crust, such as breaking
  strain and shear modulus, have implications for the detection of
  gravitational waves from a neutron star as well as bursts from Soft
  Gamma-ray Repeaters (SGRs). These properties are calculated here for
  three different crustal compositions for a non-accreting neutron
  star that results from three different cooling histories, as well as
  for a pure iron crust. A simple shear is simulated using molecular
  dynamics to the crustal compositions by deforming the simulation
  box. The breaking strain and shear modulus are found to be similar
  in the four cases, with a breaking strain of ${\sim 0.1}$ and a
  shear modulus of ${\sim 10^{30}\;\mathrm{dyne\,cm^{-2}}}$ at a
  density of $\rho = 10^{14}\,\mathrm{g\; cm^{-3}}$ for simulations
  with an initially perfect BCC lattice. With these crustal properties
  and the observed properties of {PSR~J2124-3358} the predicted strain
  amplitude of gravitational waves for a maximally deformed crust is
  found to be greater than the observational upper limits from
  LIGO. This suggests that the neutron star crust in this case may not
  be maximally deformed or it may not have a perfect BCC lattice
  structure. The implications of the calculated crustal properties of
  bursts from SGRs are also explored. The mechanical properties found
  for a perfect BCC lattice structure find that crustal events alone
  can not be ruled out for triggering the energy in SGR bursts.
\end{abstract} 


\section{Introduction}

The strength of the material that comprises neutron-star crusts determines the maximal strength of gravitational radiation from a single deformed neutron star \citep{chamel08}. Furthermore, yielding events of the neutron star crust have also been associated with bursts from Soft Gamma-ray Repeaters (SGRs) \citep{chamel08}. The neutron star crust consists of the outer regions of the star ranging from a density of $10^6\mathrm{g \, cm^{-3}}$ up to nuclear density of $2\times10^{14}\mathrm{g\, cm^{-3}}$ \citep{lattimer04}. The less dense regions of the crust consist of a lattice of nuclei, with the ground state of a body centred cubic (BCC) lattice \citep{chamel08}. At higher densities the crust is composed of a mixture of nuclei, electrons, protons and free neutrons. The mechanical properties of the neutron star crust which are important for the consideration of crustal deformation include the breaking strain, or maximal degree of deformation before yielding, the amount of stress associated with this strain, as well as the shear modulus of crustal material.

Initial estimates of the breaking strain of the neutron star crust were made by comparing to terrestrial matter. Arguing that the impurities expected in the neutron star crust would result in weakening the structure, the breaking strain of the crust could range between $\phi_m = 10^{-5}$ and $\phi_m = 10^{-2}$ \citep{smoluchowski70}. The shear modulus of the crust has been predicted by treating the neutron star crust as a Coulomb lattice, which has the approximate relationship of $\mu \propto (Ze)^2n^{4/3}$, where the charge $Z$ ranges between 30 and 50, and $n$ is the number density, which leads to a expected shear modulus of $\mu = 10^{30} \mathrm{dyne\, cm}^{-2}$ \citep{smoluchowski70}.

Recently, with molecular-dynamics simulations, the breaking strain of a system representing an accreted neutron star crust has been calculated. The crustal composition of the accreted crust consists of isotopes ranging from $Z=8$ to $Z = 47$ \citep{gupta07}. The breaking strain for the accreted crust was found to occur around $\phi_m = 0.1$, which is larger than had been predicted, and this result was also found to only moderately affected by the introduction of impurities, defects and grain boundaries \citep{horowitz09b}.

In this work molecular-dynamics simulations are used in order to investigate the crustal properties of a non-accreting neutron star. The molecular-dynamics calculations are carried out with the open source software LAMMPS (Large Atomic/Molecular Massively Parallel Simulator) \citep{plimpton95}, which is available from Sandia Laboratories\footnote{http://lammps.sandia.gov}. The breaking strain and shear modulus found with these simulations are used in order to place limits on the strain amplitude for a gravitational wave signal expected for a fully deformed neutron star. Limits are also placed on the fracture size required for the energy in a corresponding SGR burst.  In the following section the implementation and  parameters of the molecular-dynamics simulations are discussed. In Section~\ref{sec:results} the results of these simulations are presented.  The simulation results are applied to the emission of gravitational waves and SGR bursts in section~\ref{sec:discussion}. In Section~\ref{sec:conclusions} we summarize our findings. 

\section{Composition and Molecular Dynamics}

In order to calculate the composition of a non-accreting neutron star crust we used the reaction rate network {\tt torch}\footnote{http://cococubed.asu.edu/code\_pages/codes.html}\citep{timmesapjs00}.  The {\tt torch} software calculated the abundances of 489 different isotopes as the neutron star material cooled at densities ranging from $10^{6} - 10^{11}\;\mathrm{g\, cm^{-3}}$. The compositions were calculated for three different cooling curves -- cooling via modified Urca, cooling appropriate for a thick crust \citep{latt94}, as well as a thin crust. Each of the cooling scenarios resulted in a different crustal composition,  see \citet{hoffman09} for more details.

Out of the total 489 different isotopes, only the top three most abundant isotopes were used in the molecular dynamics simulations.  For the modified Urca cooling case the top three isotopes were found to be $^{56}$Fe, $^{54}$Fe,  and $^{60}$Ni. In the case of neutron star cooling with a thick crust the top three isotopes were $^{56}$Fe, $^{60}$Ni,  and $^{52}$Cr. In the thin crust case $^{54}$Fe, $^{58}$Ni,  and $^{56}$Fe were found to have the top abundances. Table \ref{tab:isotopes} summarizes these top three isotopes, as well as the rescaled mass fractions in order for the composition to sum to unity. The table also includes the number fraction of each isotope used in the simulations and the properties of each of the isotopes. 

\begin{table}
\begin{center}
\begin{tabular}{lcccccc}\hline
Isotope & A & Z & Mass  & Rescaled  & Number & Mass\\ 
              &     &     &  Fraction  &       X                          &  Fraction & \\ \hline
\multicolumn{7}{c}{Modified Urca Composition} \\ \hline
$^{56}$Fe & 56 & 26 & 0.559 & 0.6014 & 0.5961 & 1.0 \\
$^{54}$Fe & 54 & 26 & 0.3187 & 0.3429 & 0.3524 & 0.964\\ 
$^{60}$Ni & 60& 28 & 0.05175 & 0.0557 & 0.0515 & 1.071\\ \hline
\multicolumn{7}{c}{Thick Crust Composition} \\ \hline
$^{56}$Fe & 56 & 26 & 0.9286 & 0.9429 & 0.9431 & 1.0 \\
$^{60}$Ni & 60 & 28 & 0.03138 & 0.0319 & 0.0298 & 1.071\\ 
$^{52}$Cr & 52 & 24 & 0.02485 & 0.0252 & 0.0271 & 0.929\\ \hline
\multicolumn{7}{c}{Thin Crust Composition} \\ \hline
$^{54}$Fe & 54 & 26 & 0.6477 & 0.6675 & 0.6808 & 0.964 \\
$^{58}$Ni & 58 & 28 & 0.2209 & 0.2276 & 0.2161 & 1.036\\ 
$^{56}$Fe & 56 & 26 & 0.1018 & 0.1049 & 0.1031 & 1.0\\ \hline
\end{tabular}
\caption[Isotope number fractions]{\label{tab:isotopes} The top three isotopes from the {\tt torch} calculations for three different cooling scenarios: modified Urca, thick crust, and thin crust cooling. The original mass fraction from the {\tt torch} calculations, as well as the rescaled mass fraction (X) are included. The number fractions determine the number of each type of isotope used in the molecular dynamics simulations. The mass value is the scaled simulation parameter, where in the simulations the mass is unity for $A = 56$.}
\end{center}
\end{table}

The mechanical properties of the neutron star crustal material are investigated by applying a shear to the material. The shear is simulated using molecular dynamics.  Molecular dynamics is the integration of Newton's equations of motion for a system of particles. The molecular dynamics simulations calculate the effect of a specified force applied collectively to a system. Within the simulations the position, velocity, and acceleration of each particle are calculated, leading to a determination of the thermodynamic properties of the system. For the systems of interest in this work the particles interact via a Yukawa, or screened Coulomb, potential 
\begin{equation} \label{eq:yuk}
V_{ij} = \frac{Z_iZ_ke^2}{r}{\rm e}^{-\kappa r}, 
\end{equation}
where $r$ is the distance between the particle pair, $Z$ is the particle charge, and $\kappa$ is the inverse screening length. The inverse screening length is given as $\kappa = 2\,\alpha^{1/2}k_F/\pi^{1/2}$, where $k_F = (3\pi^2n_e)^{1/3}$, with $n_e=\langle Z \rangle n$ as the electron number density.  The molecular dynamics simulations are conducted using LAMMPS, with the evolution of the particle motion calculated with velocity-Verlet integration \citep{vverlet}. The velocity-Verlet algorithm is a variation of Verlet integration \citep{verletref}. It is a second-order integration method which is time reversible and is a result of the expansion of the particle position with time. At a given timestep the velocity-Verlet algorithm calculates both the velocity and position of each particle, as well as the force of each of the interactions. In order to decrease the calculation time, LAMMPS makes use of parallel computing \citep{plimpton95} and neighbour lists. The neighbour lists each contain a list of particles within a specified distance from each other and only particle pairs within a specified cut-off radius are used in the interaction calculations. LAMMPS version 10 Jul 2010 was used in the simulations discussed here. 

The simulation parameters are expressed in reduced units based on the Yukawa potential (Eq.~\ref{eq:yuk}). These reduced units introduce a set of characteristic quantities: length, energy, mass, and charge. The characteristic length scale, $a$ is  $a = n^{-1/3}$, where $n$ is the number density. The characteristic energy scale is defined as 
\begin{equation}
U_a = \frac{(Z e)^2}{a},
\end{equation}
which is expressed with respect to a charge of $Z=26$. The characteristic mass, $m_a$, is expressed in units of the mass of iron, or $m_a = A/56$, thus for an iron ion $m_a$ is unity. In a similar manner, the characteristic charge is expressed in units of the iron charge, thus $Z_a = Z/26$. These characteristic quantities are used in order to relate reduced unit simulation parameters to the corresponding physical quantities. For example the characteristic quantities of $^{56}$Fe at a density of $\rho = 10^{11}\mathrm{g\; cm^{-3}}$ are a characteristic length of $a \sim 10^{-11}$cm and a characteristic energy of $U_a \sim 10^{-5}$erg. The relation between the reduced and physical values are listed in Table \ref{tab:units}. 

\begin{table}
\begin{center}
\begin{tabular}{ll}\hline
 Unit & Reduced Unit\\ \hline
Temperature & $T^* = \dfrac{T k_b}{U_a}$ \\ 
Energy & $E^* = \dfrac{E }{U_a}$ \\ 
Distance & $r^*=\dfrac{r}{a}$ \\
Pressure & $P^* = P \dfrac{a^3}{U_a}$ \\ 
Time & $t^* = t \left(\dfrac{2U_a}{ma^2}\right)^{1/2}$ \\ \hline
\end{tabular}
\caption[Unit conversions for the reduced units]{\label{tab:units} The reduced unit system, based on the Yukawa potential, used in the {\tt LAMMPS} simulations. Parameter values quoted from the simulations are in reduced units. The physical quantities are determined by using appropriate values the characteristic distance, $a$, and energy, $U_a$, parameters for the material.}
\end{center}
\end{table}

Within all the simulations the the number density is set to unity. Periodic boundary conditions are also used with the shear introduced by deforming the box in the x-direction. The degree of strain is determined by taking the ratio of the displacement of the top of the simulation box to the original box length, $\Delta x/ l$.  The strain rate is the the velocity of the shear applied at the top of the box divided by the length of the box, $v/l$.

The simulations all begin with each of the particles given an initial velocity kick, which gives the temperature of the simulation below the melting temperature. After 200 steps of equilibrating the particles to a specified temperature the box is deformed, holding the temperature constant through temperature rescaling. The velocity-Verlet integration, which calculates the positions and particle velocities, is performed on a system of particles which is consistent with a microcanonical ensemble, constant number, volume and energy (NVE).  A nearly constant temperature is maintained by rescaling the velocities.  A system of particles consistent with a canonical ensemble, constant number, volume, and temperature (NVT), could have also been selected and used for temperature control, but the velocity-rescaling thermostat controls the temperature in a more gentle and direct manner than the Nos\'e-Hoover or Langevin thermostat.  Of course, the latter would better approximate the canonical ensemble (NVT); however, for simplicity and for comparison with previous work \citep[e.g.][]{horowitz09b}, we also used velocity rescaling.  The interaction calculation is cut-off at a distance of $r_c  = 8/\kappa$ between particle pairs, where $\kappa$ is the inverse screening length. The pair-coefficients of the different particle pair interactions, $Z_iZ_je^2$, are listed in Table \ref{tab:pair}.  The average mass, average charge, inverse screening length, cut-off radius, and impurity factor of the composition are listed in Table \ref{tab:simp} for the three impure cases. 

\begin{table}
\begin{center}
\begin{tabular}{l | ccccc}\hline
& $^{56}$Fe & $^{52}$Cr & $^{54}$Fe  & $^{58}$Ni & $^{60}$Ni \\ \hline
$^{56}$Fe & 1.0 & 0.9231 & 1.0 & 1.0769 & 1.0769 \\
$^{52}$Cr & -- & 0.8521 & 0.9231 & 0.9941 & 0.9941 \\
$^{54}$Fe & -- & -- & 1.0 & 1.0769 & 1.0769 \\
$^{58}$Ni & -- & -- & -- & 1.1598 & 1.1598 \\
$^{60}$Ni & -- & -- &  -- & -- & 1.1898 \\ \hline
\end{tabular}
\caption{\label{tab:pair} The pair-coefficients, $Z_iZ_j\,e^2$, for the different particle pairs. The pair-coefficients are used in the calculation of the interaction potential. These pair-coefficients are used in the four different cases of crustal simulations, the three impure cases and a pure iron crust case.}
\end{center}
\end{table}

\begin{table}
\begin{center}
\begin{tabular}{lccccc}\hline
& $\langle A \rangle $ & $ \langle Z \rangle$ & $\kappa$ & r$_c$ & $Q_\mathrm{imp}$ \\ \hline
Mod. Urca & 55.501 & 26.103 & 0.8847 & 9.043 & 0.195 \\
Thick & 56.011 & 26.005 & 0.8836 & 9.054 & 0.228 \\
Thin & 55.071 & 26.432 & 0.8884 & 9.005 & 0.678 \\ \hline
\end{tabular}
 \caption[Parameters for the impure simulations]{\label{tab:simp} The inverse screening length and the corresponding cut-off radius used in the simulations. The parameters $\langle A \rangle $ and $ \langle Z \rangle$ were used in the calculation simulation parameters. The $Q_\mathrm{imp}$ value, where $Q_\mathrm{imp} = \langle Z^2 \rangle - \langle Z \rangle ^2$ \citep{itoh93},  indicates the degree of impurity of the sample for the three impure crustal cases.}
\end{center}
\end{table}

\section{Results}
\label{sec:results}

The initial molecular dynamics simulations performed included tests to examine the effect of the size of the simulations as well as varying the strain rate in order to look for convergence of the material properties examined. In order for the shearing simulations to be run below the melting temperature,  the melting temperature of the various materials was also determined. With these initial test simulations, parameters for the simple shear simulations were determined. The simple shear simulations were all performed at a temperature below the melting temperature of the crystal. A simulation box size of $25\times25\times25$ unit cells for simple shear simulations of 31250 particles. Strain rates of 20$\times$, 5$\times$, and 2.5$\times10^{-6}$ were examined in the simple shear calculations. The application of the shear was also used on crystals with an imperfect crystal structure and compared to the results of a perfect BCC crystal lattice structure. The temperature dependence of the stress-strain relationship was also examined for three different temperature regimes. The possibility of a second yielding event was investigated via simulations run to a larger degree of strain. Finally, the reversibility of the yielding event, due to the applied shear, was also examined.

\subsection{Test Simulations}
\label{sec:test-simulations}
The test simulations included determining the melting temperature of the different compositions as well as investigating  size effects. The melting temperature was determined in order to ensure the simple shear simulations were below the melting point of the compositions. The box size tests were performed in order to determine the effect the box size had on the simulation, as well as finding the smallest number of particles which could be used to gain a reliable result from the simulations. 

\subsubsection{Melting Temperature}\label{sec:melting-temperature}

The melting temperature of the three impure crustal compositions, as well as the pure iron case, was determined by increasing the temperature of the simulation box. In these simulations, unlike the shearing simulations, the temperature was controlled by a Langevin thermostat. With a Langevin thermostat the system of particles is within a heat bath and the temperature is controlled through a frictional force as well as a random force, which imparts a random velocity kick \citep{langevin}. The melting temperature was defined as the temperature in which the mean squared displacement of the particles has  the greatest change in the value between time steps. The simulation boxes were heated from a temperature of  $T^* = 0.001$ to a value of $T^* = 0.02$ over  $10^6$ steps. All four cases were found to melt at around the same temperature of  $T^*{\sim}0.01$. To compare this to a physical temperature, at a density of   $\rho = 10^{11}\,\rm{g/cm^3}$ these melting temperatures correspond to ${\sim}\, 7 \times 10^{8}$\,K. The temperature of the simple shear simulations are always less than this melting temperature.  

\subsubsection{Size Effects}
\begin{figure}
\includegraphics[width = 3.3in]{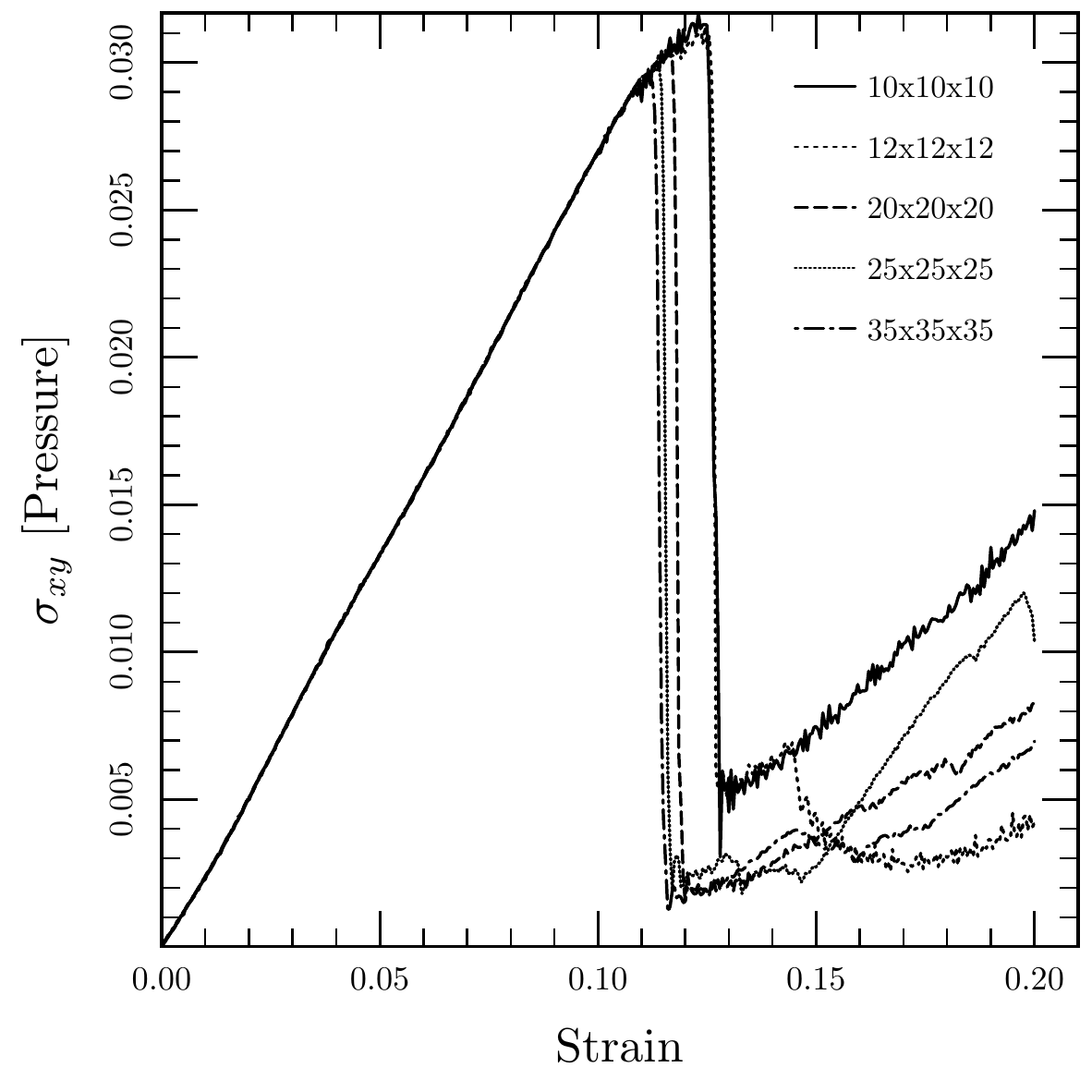}
\caption[Breaking strain and the size effects of a BCC crystal]{The stress-strain relationship of a pure BCC crystal comparing five different sizes of simulations. The legend indicates the unit cell volume of the simulation box. The number of particles range from 2000 in the smallest simulation box to 85750 particles in the largest simulation box size.   For the same applied strain rate of $\dot{s} = 20\times10^{-6}$ the five sizes of simulations share the same shear modulus, but the breaking strain is different. The larger simulation boxes are found to yield before the smaller simulation boxes.}
\label{fig:bccsize}
\end{figure}

In order to examine the possible simulation size effects a BCC iron lattice with periodic boundary conditions was strained at a rate of $20\times10^{-6}$ for five different box sizes. The simulation box sizes included boxes with unit cell volumes of  $10\times10\times10$,  $12\times12\times12$, $20\times20\times20$,  $25\times25\times25$, and $35\times35\times35$, corresponding to 2000, 3456, 16000, 31250, and 85750 particles, respectively. The result of deforming these five different simulation box sizes are compared in Figure \ref{fig:bccsize}, which displays an apparent difference in the box size and yielding stress and strain. The shear modulus is the same in all five cases, but simulations with a larger number of particles are found to yield earlier. The stress-strain relationship for the $35\times35\times35$ and $25\times25\times25$ unit class box sizes are similar, and in order to decrease the calculation time a $25\times25\times25$ box size is used for simulations of the different crustal compositions.

\subsection{Simple Shear}
\begin{figure}
\includegraphics[width=3.3in]{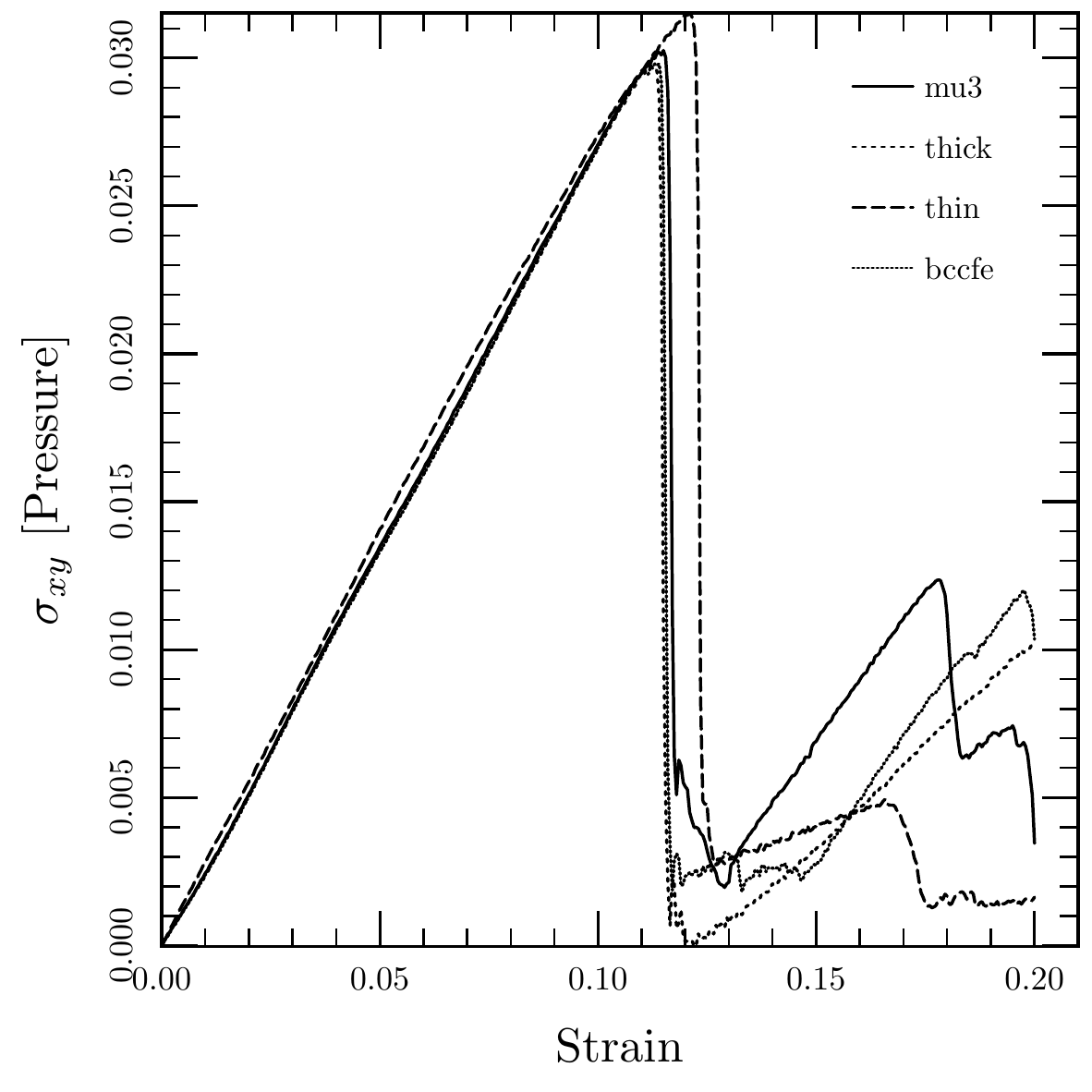}
\caption{\label{fig:sscomp}The stress-strain relationship for the three impure crustal compositions: modified Urca (mu3), thick crust (thick), and thin crust (thin), compared to a pure iron BCC lattice (bccfe). A strain rate of $20\times10^{-6}$ was used in  all four of these simulations. The shear modulus was found to be  similar for all four of the simulations. The thin crust yielded to the deformation at a higher strain than the other three simulations. As these results use a perfect BCC lattice structure,  these simulations set an upper limit for the shear modulus and breaking strain of a neutron star crust.}
\label{fig:ev20ss}
\end{figure}

The four cases of three impure compositions appropriate for cooling by modified Urca, a thick crust, and a thin crust, as well as a pure iron crust were deformed in molecular dynamics simulations in order to determine the mechanical properties. The shear was applied by deforming the simulation box in the x-direction. Three different strain strain rates were applied in the simulations, 20, 5, and 2.5$\times10^{-6}$, in order to test for convergence of the mechanical properties. The simulation boxes  had periodic boundary conditions and contained 31250 particles in all cases. Each simulation stared in a perfect BCC lattice configuration, for the impure cases the isotopes were randomly assigned to lattice locations. The stress-strain relationships at a stain rate of $20\times10^{-6}$ for the four cases are compared in Figure \ref{fig:sscomp}. The yielding strain and shear modulus determined for the four cases at the three different strain rates are listed in Table \ref{tab:bsall}. The pure iron, modified Urca and thick crust compositions share very similar breaking strains, of $\phi_m \sim 0.11$, and shear moduli, of $\mu^* \sim 0.21$. The thin crust composition is found to yield to the applied stress at a later degree of deformation than the other three compositions, at around a breaking strain of  $\phi_m \sim 0.12$. These results of a breaking strain of ${\sim}0.1$ for  non-accreting crustal compositions is ten times larger than the initial estimates from \citet{smoluchowski70}, but the results are consistent with previous molecular dynamics simulations for an accreted crustal composition \citep[see][]{horowitz09b}. 

\begin{table}
\begin{center}
\begin{tabular}{cccc} \hline
Composition & Strain Rate & Yielding & Shear \\
	& ($\times10^{-6}$)  & Strain & Modulus \\ \hline
Pure & 20 & 0.114508 & 0.27217 \\
Iron & 5 & 0.112896 & 0.27219 \\
   & 2.5 & 0.111295 & 0.27220 \\ \hline
Modified & 20 & 0.115620 & 0.2735 \\
Urca  & 5 & 0.113396 & 0.2736 \\
	& 2.5 & 0.113174 & 0.2735 \\ \hline
Thick  & 20 & 0.113285 & 0.27179 \\
Crust  & 5 & 0.111895 & 0.27178 \\
	& 2.5 & 0.111673 & 0.27179 \\ \hline
Thin & 20 & 0.122846 & 0.26644 \\
Crust  & 5 & 0.120122 & 0.27620 \\
	& 2.5 & 0.119455 & 0.27711 \\
\end{tabular}
\caption{\label{tab:bsall} A comparison of the the yielding strains and shear moduli for the four different compositions. The ratio of the strain rate to the plasma frequency is  56.42, 3.979, and 1.989$\times10^{-7}$, for strain rates of 20, 5, and 2.5$\times10^{-6}$, respectively.} 
\end{center}
\end{table}

\subsection{Imperfect Crystal Structure}
\label{sec:imperf-cryst-struct}
\begin{figure}
\includegraphics[width = 3.3in]{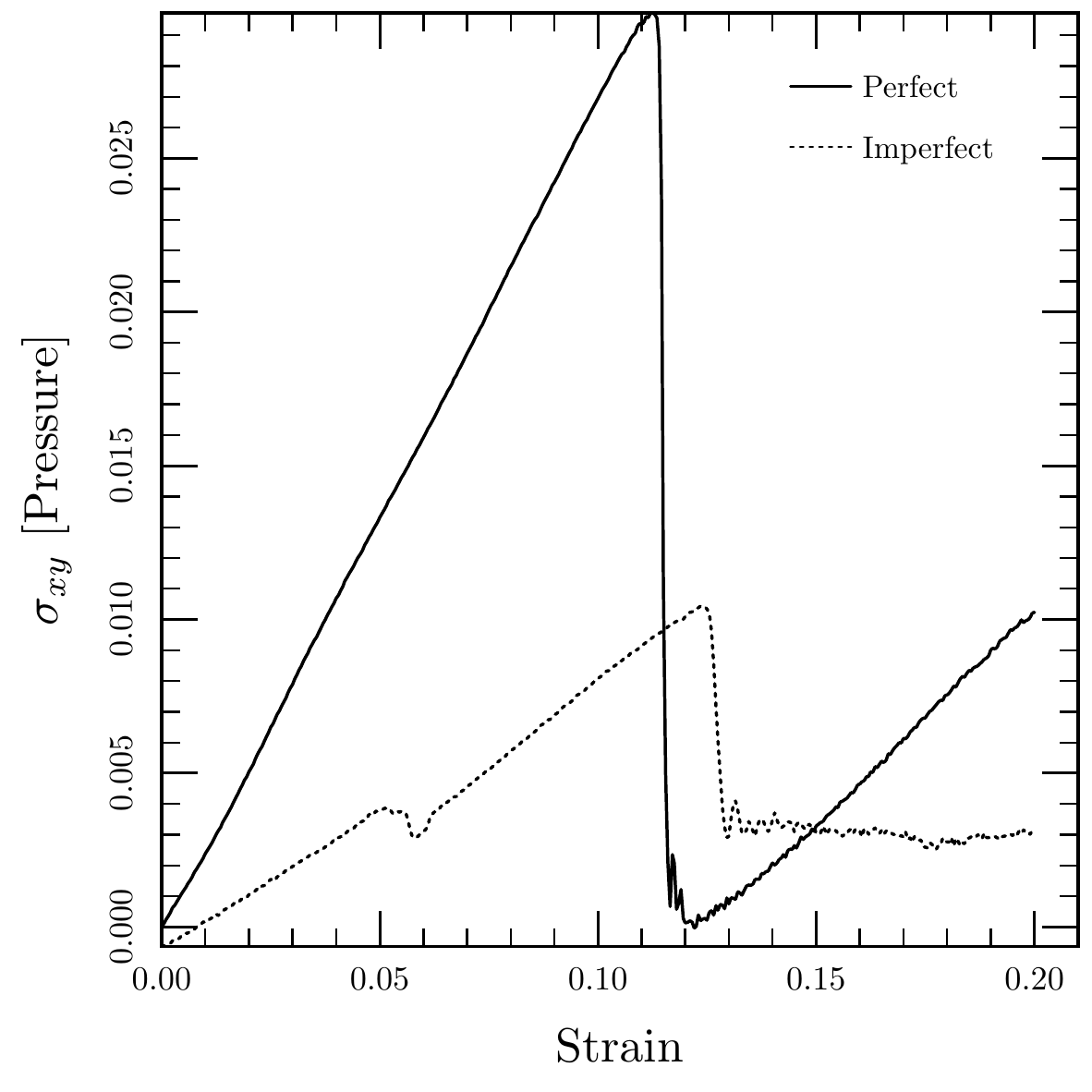}
\caption{A comparison of the stress-strain relationships of a perfect and an imperfect crystal which contains defects for the case of a thick crust composition. A strain rate of  $\dot{s} = 20\times10^{-6}$ was applied in the X-direction for both simulations. The two crystals display differing  behaviours in their stress-strain relationships.}
\label{fig:perimp}
\end{figure}

The shearing simulations discussed above initially had a perfect BCC lattice structure. The addition of defects to the simulation lattice would be expected to decrease the crystal strength, as a result the perfect BCC lattice structure simulations may represent an upper limit on the crustal strength. It should be noted that previous molecular dynamics simulations which included defects were found to only moderately affect the strength \citep{horowitz09b}. In the case of a non-accreting neutron star crustal composition the effect of defects are examined by melting the perfect BCC lattice structure and then allowing the simulation box to recrystallize. The application of a shear to a perfect BCC lattice and an imperfect lattice is compared in Figure \ref{fig:perimp} for a thick crust composition. The behaviour, as displayed in Figure \ref{fig:perimp}, of the two different initial crystal types have different behaviours in terms of the stress-strain relationship and the resulting shear modulus and the yielding strain. The difference in the effect of defects as reported here as opposed to \citet{horowitz09b} could be due to the number of added defects or grain boundaries. 

\subsection{Temperature Dependence}
\label{sec:temp-depend}
\begin{figure}
\includegraphics[width = 3.5in]{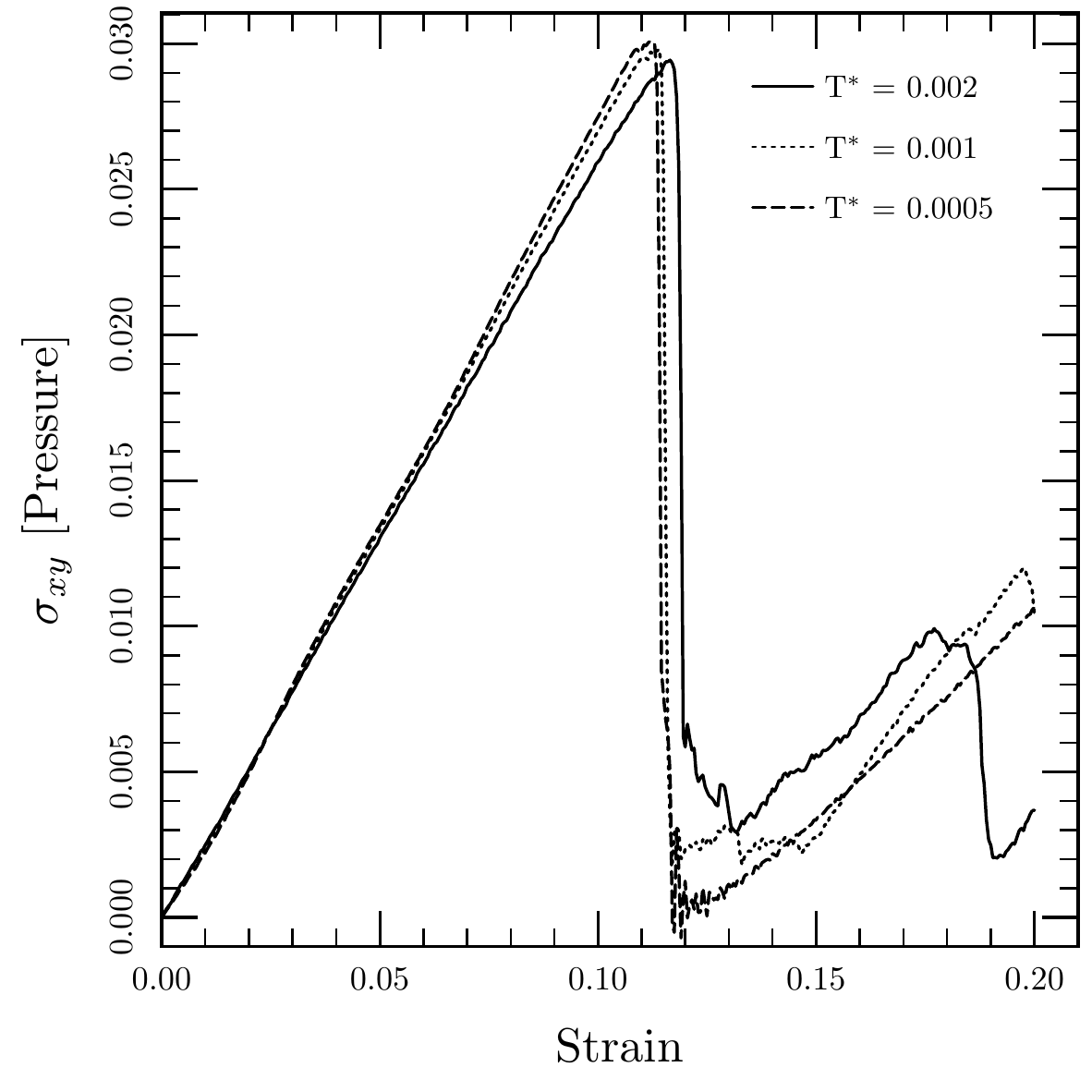}
\caption{The stress-strain relationship for pure iron BCC crystals at three different temperatures.  At $T^*{\sim0.01}$ the crystal melts; all simulation temperatures are below the melting temperature. A strain rate of $20\times10^{-6}$ was applied in all three cases. For each temperature the shear modulus and breaking strain slightly differ, though $T^* = 0.001$ and $T^*= 0.0005$ were found to be very similar. The shear modulus of the colder simulation was slightly larger then the warmer simulations. The colder simulation is found to yield at a smaller strain, but higher stress, than the hotter simulations. }
\label{fig:bcctemp}
\end{figure}

The temperature dependence of the stress-strain relationship was investigated by performing additional simulations at  temperatures twice and half as hot as the original simulation, for a total of three different temperatures examined. Thus, for each crustal composition, temperatures of $T^*=0.002$, $T^*=0.001$, and $T^*=0.0005$, which correspond to physical temperatures of $6.9\times10^8\;$K, $3.5\times10^8\;$K, and $1.7\times10^8\;$K at a density of $\rho = 10^{14}\,\mathrm{g\,cm^{-3}}$, were investigated. A strain rate of $20\times20^{-6}$ was applied to an initially perfect BCC lattice structure in all cases. The three temperature stress-strain relationship results for a pure iron perfect BCC lattice are compared in Figure \ref{fig:bcctemp}. Though the relationships are very similar, the shear modulus for the colder simulations is slightly larger than the hotter temperatures and the material yields to applied stress earlier in the colder simulations as opposed to the hotter. This behaviour is also observed in the three impure crustal compositions, with the shear modulus and the breaking strains listed in Table \ref{tab:temp} for the four cases investigated. The temperature dependence observed in these simulations is expected from models of shear modulus-temperature relationships,  e.g. \citet{npref}, where the shear modulus is found to  decrease with increasing temperature. 

\begin{table}
\begin{center}
\begin{tabular}{cccc}\hline
Crust & Temperature & Breaking & Shear  \\ 
Composition & $T^*$  &Strain & Modulus  \\ \hline
 Pure & 0.002 & 0.11712 & 0.26069  \\
Iron & 0.001 & 0.114508 & 0.27217  \\
& 0.0005 & 0.112285 & 0.27821  \\ \hline
Modified & 0.002 & 0.116064   &  0.26253  \\
 Urca& 0.001 & 0.115620 & 0.27351   \\
 &  0.0005 & 0.112952 &  0.29558   \\ \hline
Thick & 0.002 & 0.115397 & 0.26052  \\
Crust & 0.001 & 0.113285 & 0.27179  \\
& 0.0005 & 0.112396 & 0.27789  \\ \hline
Thin & 0.002 & 0.122846 & 0.26644  \\
Crust & 0.001 & 0.121901 & 0.27637  \\
& 0.0005 & 0.122401 & 0.28060  \\ \hline
\end{tabular}
\caption{\label{tab:temp} The temperature, breaking strain and shear modulus for the three different impure crustal compositions, as well as the pure iron case. The three different crust compositions melt at approximately $T^*=0.01$, all temperature simulations are below the melting temperature. At a density of $\rho = 10^{14}\,\rm{g/cm^3}$ the corresponding physical temperatures are $6.9\times10^{8}$\, K for $T^* = 0.002$, $3.5\times10^{8}$\, K for $T^* = 0.001$ and $1.7\times10^{8}$\,K for $T^* = 0.0005$. }
\end{center}
\end{table}

\subsection{Second Yielding Events}
\label{sec:second-yield-events}
\begin{figure}
\includegraphics[width=3.5in]{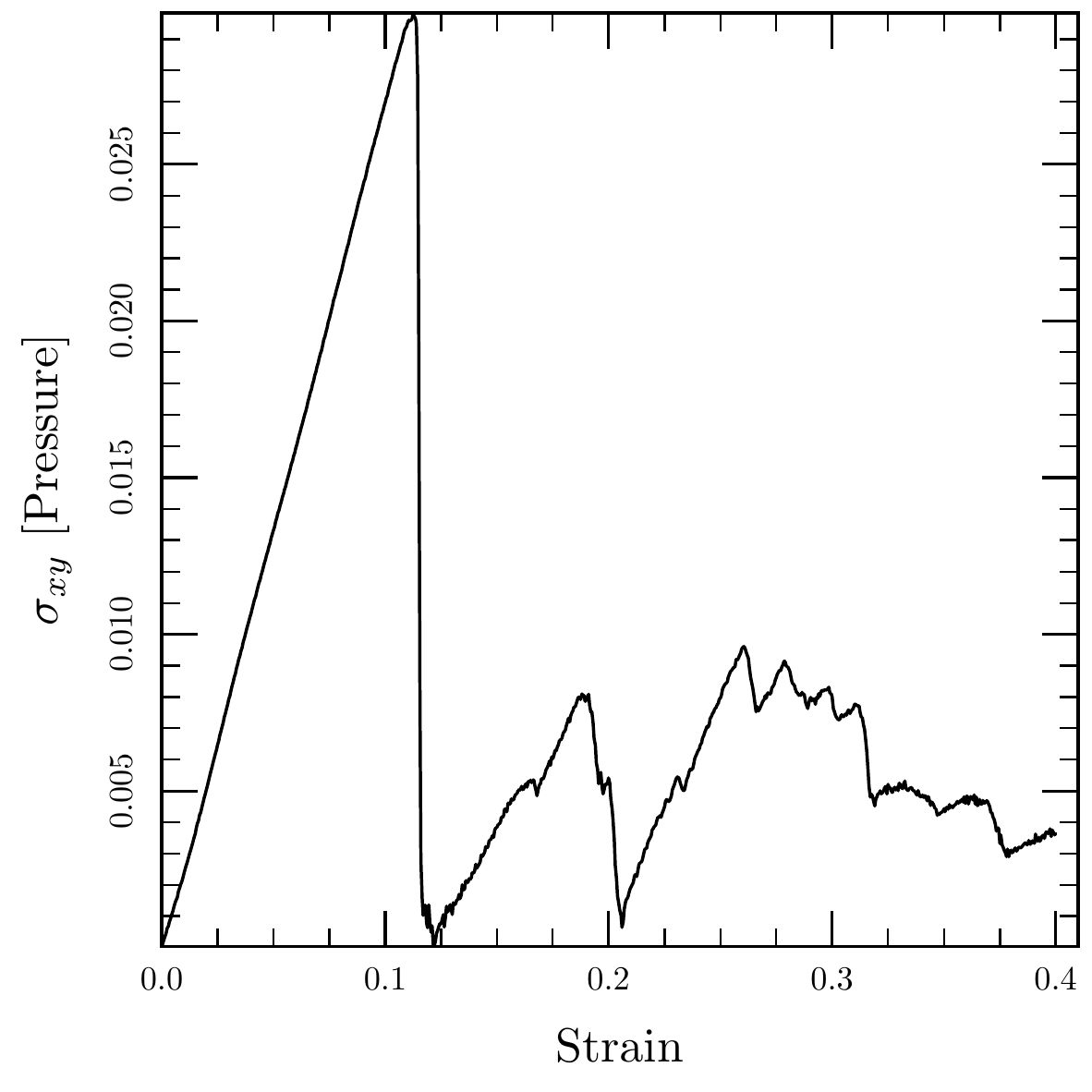}
\caption{The stress-strain relationship of a 31250 particle,  pure iron BCC crystal with an inverse screening length of $\kappa = 0.8835$ with  a strain rate of $20\times10^{-6}$ applied. The simulation box was deformed to a strain of $\Delta x/l =0.4$ in order to investigate the occurrence of a second material failure. A second yielding event occurs for the crystal  between a strain of 0.133 to 0.186 with a shear modulus of $\mu^*{ \sim 0.11}$ in reduced pressure units.}
\label{fig:2bbccss}
\end{figure}

In the lifetime of a neutron star, the crust would have likely undergone multiple yielding events. Subsequent yielding events, after the first major event, are investigated in these simulations by deforming the simulation box to a strain of 0.4, where the original simulations were deformed to a 0.2 stain.  In these longer simulations a strain rate of  $20\times10^{-6}$ was applied to all four types of crustal composition. The simulations discussed here started with a perfect BCC lattice structure.  The stress-strain relationship  for the pure iron case is displayed in Figure  \ref{fig:2bbccss}. As the stress is applied,  the crystal undergoes a linear regime before yielding to the applied stress, but after the major yielding event the stress-strain relationship undergoes an additional linear regime, which has a different slope, or shear modulus, as the initial event. In the case of the pure iron BCC crystal the shear modulus  before the initial yielding point is  $\mu^*\, {\sim}\, 0.27$ and in the second linear regime, measured between values of  a strain of 0.133 and 0.186 the shear modulus is $\mu^*\, {\sim}\, 0.11$.  In the cases with the modified Urca, thick and thin crustal compositions similar behaviour occurs with the shear modulus changing after the initial yielding point. In the case of the thick crust composition the reduced shear modulus changes from $\mu^*\, {\sim}\, 0.27$ before the first major yielding event to $\mu^*\, {\sim}\, 0.12$ for the second. Comparing to the results of the imperfect crystal in section \ref{sec:imperf-cryst-struct}, the behaviour of the second break is similar to the imperfect crystal, though the linear regime has a slightly larger slope for the initially perfect lattice structure crystal. 

\subsection{Yielding Reversibility}
\label{sec:yield-revers}
\begin{figure}
\includegraphics[width = 3.3in]{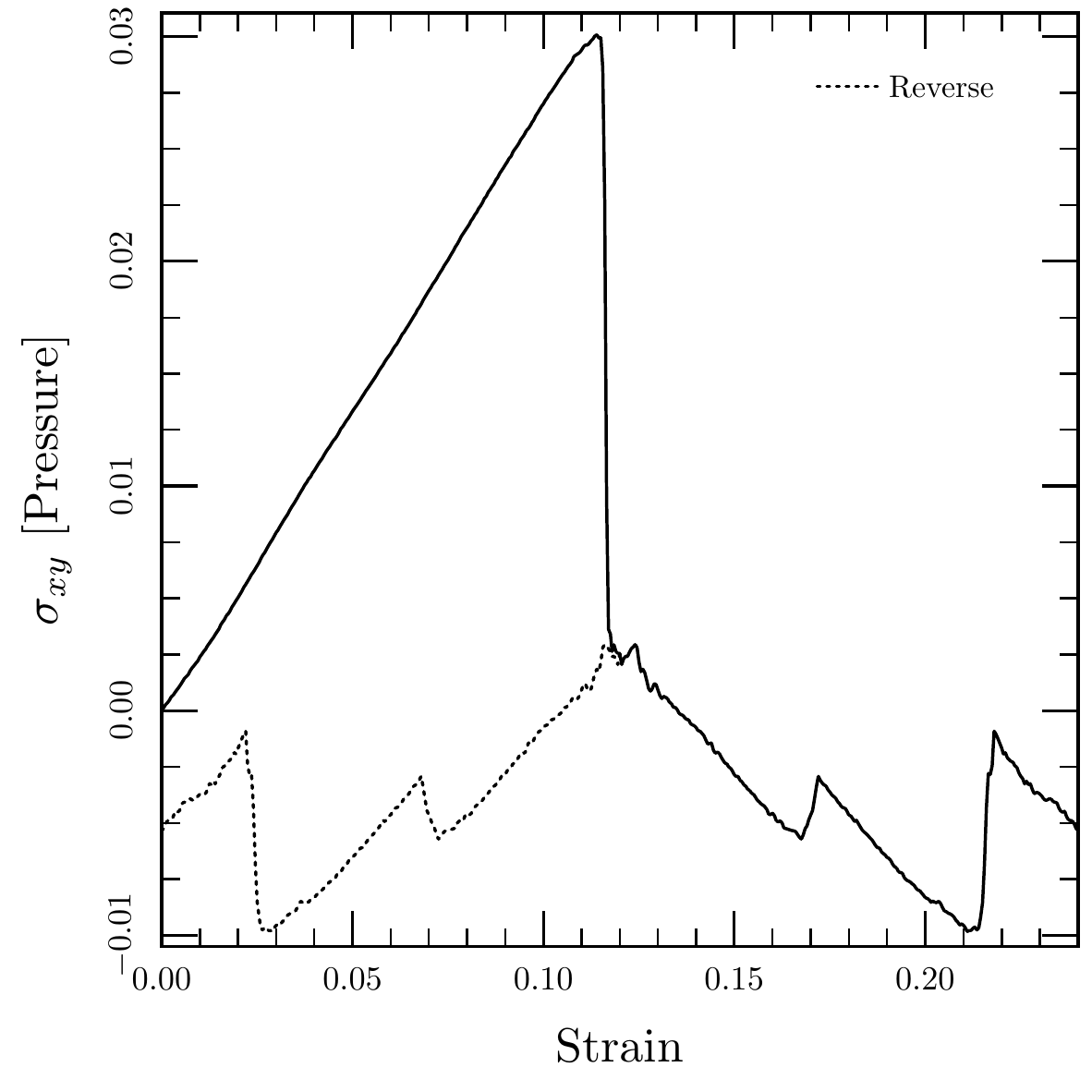}
\caption{The stress-strain relationship of a pure iron BCC crystal with an inverse screening length of $\kappa = 0.8835$ and a strain rate of $20\times10^{-6}$ applied. The deformation of the box was reversed after the material yields. For illustrative purposes the solid line indicates the strain as is the box deformation was continued in the forward direction, the dotted line indicates the box deformation as it is being deformed in the reverse direction. In the forward direction the shear modulus was $\mu^*{\sim 0.27}$, after the box deformation was reversed the shear modulus is $\mu^*{\sim 0.17}$. The shear modulus was measured between a strain of 0.175 and 0.2115.}
\label{fig:fbbccss}
\end{figure}

In order to investigate the reversibility of the yielding event, the direction of the box deformation was reversed after the major yielding event occurs. If the yielding event was reversible than the shear modulus would be the same deforming in the reverse direction as the the forward direction. In these simulations the simulation box containing a perfect BCC lattice is deformed at a strain rate of $20\times10^{-6}$, upon yielding the box is then deformed in the opposite direction, but at the same strain rate. The stress-strain relationship for this type of simulation is displayed in Figure \ref{fig:fbbccss} for a pure iron perfect BCC lattice structure; for plotting purposes the amount of strain in the figure continues to increase after reversing the shear direction as (step size$\times \Delta t \times 20\times10^{-6}$).  In this case of the pure Iron BCC crustal the shear modulus is $\mu^*\,{ \sim}\, 0.27$ in the forward direction. After reversing the direction of the shear direction the shear modulus is measured to be $\mu^*\,{ \sim}\, 0.17$, measured between a strain of 0.175 and 0.2115 on the corresponding figure. As the shear modulus does not return to the original value, the yielding event is not reversible, similar results are found for the three impure composition cases.  The reduced shear modulus in the reverse direction, $\mu^*\,{ \sim}\, 0.17$, is larger than what was found for the second break, $\mu^*\,{ \sim}\, 0.11$ in the pure iron crystal case. Consequently, the shear modulus for the reverse direction would also be expected to be greater than for the imperfect crystal. 

\section{Discussion}
\label{sec:discussion}
The simple shear simulations of both the non-accreted crustal compositions, presented above, and the accreted crustal composition \citep[see][]{horowitz09b} find that the neutron star crust yields at a strain of ${\sim}0.1$, indicating that the neutron star crust is much stronger than the initial estimates. The results of a stronger crust could have implications for the observational consequences of crustal yielding events. Comparing the results of the pure iron case to the simulation results which contained three different isotopes did not indicate much of a change in the material properties for simulations with impurities and those with out. The addition of imperfections, while the system may globally have a BCC structure, indicate a difference in the stress-strain relationship as compared to systems which started in a perfect  BCC lattice structure. Depending on the strength of the neutron star crust, detectable gravitational waves may be emitted due to mountains supported on the surface. Yielding events in the crust have  been associated observationally with bursts from SGRs. The results of the shearing simulations on a perfect BCC lattice structure are now applied to these two observational consequences of the mechanical properties of the neutron star crust. 

The strain at which the crustal material was found to yields was at ${\sim}0.1$, and the shear modulus from the simulations was found to be $\mu^*{\sim 0.27}$  in reduced pressure units. To apply these results to gravitational waves and SGR bursts the simulation parameters need to be converted to physical units. The reduced pressure, or the pressure from the simulations, is expressed as:
\begin{equation}
P^* = P\frac{a^3}{U_a}, 
\end{equation}
where $P$ is the physical pressure, $U_a$ is the characteristic energy, and $a$ is the characteristic distance. The observational consequences of crustal cracking are typically investigated at the bottom of the neutron star crust, at a density of $\rho = 10^{14}\mathrm{g\; cm^{-3}}$. These simulation results are only appropriate for densities closer to the neutron star surface, but in order to compare to previous work, the results from the simulations are extrapolated to the greater density by taking into account the neutron fraction. At the bottom of the crust the neutron fraction is $X_n=0.8$, which is higher than at lower densities closer to the neutron star surface. The charge can be set to $Z = 20$ and the atomic mass to $A = 88$ at the bottom of the crust \citep{gregu}.  With these considerations the number density is related to the neutron fraction as: 
\begin{equation}
n = a^{-3} = \frac{\rho}{m} (1-X_n) = \frac{10^{14}\,\rm{g/cm^3}}{88\, m_u}\times 0.2,
\end{equation}
where $m_u$ is the atomic mass unit. 

At a density of $\rho = 10^{14}\,\mathrm{g\,cm^{-3}}$, the bottom of the neutron star crust, the characteristic length is  $a = n^{-1/3} = 1.94\times 10^{-12}\,\rm{cm}$ or 19.4\,fm and the characteristic energy, where the charge is $Z=20$, is  $U_a = (Ze)^2/a = 4.76\times10^{-5}\,\rm{erg}$, which corresponds to  ${\sim}30$\,MeV. With these characteristic values, the shear modulus of $\mu^*=0.27$ from the simulations corresponds to $\mu = 1.76\times10^{30}\,\mathrm{dyne\,cm^{-2}}$ at the bottom of the crust. This value is close to  the perviously used value of  $10^{30}\,\mathrm{dyne\,cm^{-2}}$ \citep[for example][]{ruderman69}. Note that at a density of $\rho = 10^{7}\,\mathrm{g\,cm^{-3}}$ the shear modulus would be $\mu \sim 10^{22}\,\mathrm{dyne\,cm^{-2}}$.  The strain at which yielding occurs is a unitless quantity, thus the simulation values are the physical values with a  breaking strain of ${\phi_m \sim 0.1}$ used in order to  compare the simulation results to observations. Note that the breaking strain found for the non-accreted neutron star crust is in agreement of the breaking strain calculated in \citet{horowitz09b} of close to $\phi_m = 0.1$, for an accreted crust crustal composition. 

\subsection{Gravitational Waves}
\label{sec:gravitational-waves}
Gravitational waves may be emitted by a neutron star with varying quadrupole moment. This quadrupole moment is the result of the neutron star perturbed to be asymmetric about its rotation axis due to a mountain, or deformation, on the neutron star surface. The deformation the crust can sustain depends on its mechanical properties, such as the breaking strain. The maximum quadrupole can be expressed in terms of the breaking strain and shear modulus of the crust \citep{gregu}. In these calculations the breaking strain is expressed as $Q_{22} = \gamma\, \phi_m\, I$, where $\gamma$ is a factor which depends on the mass and radius of the neutron star, $\phi_m$ is the breaking strain, and $I$ is an integral which is directly proportional to the shear modulus, and thus $Q_{22}$ is directly proportional to the shear modulus, $\mu$. Scaling by the value of $Q_{22}$ for a shear modulus of $\mu = 10^{30}\,\rm{dyne/cm^2}$ yields the following relationship
\begin{equation}
Q_{22} = 10^{38}\,\mathrm{g\, cm^2} \left(\frac{\phi_m}{10^{-2}}\right) \left(\frac{\mu}{10^{30}\mathrm{dyne\,cm^{-2}}}\right).
\end{equation}
 For the simulation results of the perfect crystal structure with a breaking strain of $\phi_m = 0.1$ and at a density of  $\rho = 10^{14}\,\rm{g\,cm^{-3}}$, where $\mu = 1.76\times10^{30}\,\mathrm{dyne\,cm^{-2}}$, the quadrupole is ${Q_{22} = 1.76\times10^{39}\rm{g\, cm^2}}$, assuming the crust is maximally deformed.

The effect of a mountain on a neutron star is  observed through the measured strain amplitude of gravitational radiation, which is given by
\begin{equation}
h = \frac{16}{5} \left(\frac{\pi}{3}\right)^{1/3}\frac{G\; Q_{22} \; \Omega^2}{d \; c^4},
\end{equation}
where $\Omega$ is the angular frequency and $d$ is the distance to the neutron star \citep{gregu}. Observational upper limits have been placed on the gravitational wave strain amplitudes of 78 radio pulsars using the Laser Interferometer Gravitational Wave observatory (LIGO) \citep{abbott07}. A detection of gravitational waves from a neutron star, with a comparison to predicted strain amplitudes could possibly put constraints on the strength and structure of the neutron star crust. 

Of the 78 different pulsars which have had upper limits placed on their gravitational wave emission, the properties of three pulsars are examined in the context of the simulation results: PSR~J1603-7202, PSR~J2124-3358, and PSR~J0534+2200 (the Crab pulsar).  Though these pulsars have likely undergone accretion events as two of the pulsars are recycled and the third is the Crab pulsar, which is in a supernova remanent and may have accreted some of the nearby matter, the breaking strain results found for the non-accreted crust are similar to those reported for an accreted crust in \citet{horowitz09b}.  The LIGO observational limits of the gravitational wave strain amplitude, $h$, are compared to the strain amplitude for the deformation of an initially perfect BCC lattice crust which is maximally deformed in Table \ref{tab:gw}. The distance, $d$, and the spin frequency, $\nu$, of the three pulsars are also listed in the table. 

\begin{table}
\begin{center}
\begin{tabular}{ccccc}\hline
Pulsar & $d$ & $\nu$ & LIGO & Predicted \\ 
  & kpc & Hz & Max & Max  \\
   &        &       &   $\log(h)$    &   $\log(h)$ \\ \hline
PSR J1603-7202 &1.64 & 67.38 & -24.58 & -25.8 \\
PSR J2124-3358 & 0.32 & 202.71 & -24.31 & -24.1 \\
PSR J0534+2200 &2 & 29.80 & -23.51 & -26.6 \\ \hline
\end{tabular}
\caption{\label{tab:gw}The distance, $d$, spin frequency, $\nu$,  and the observational and predicted gravitational wave strain amplitude, $h$, for the three pulsars of interest. The predicted gravitational wave strain amplitudes are below the observational LIGO upper limits for two of the cases. For pulsar PSR J2124-3358 the predicted strain amplitudes indicate that if the crust was initially perfectly BCC in structure and maximally deformed, than gravitational waves should have been detected from this source. The distances of each pulsar were taken from  the ATNF Pulsar Catalogue ({\tt http://www.atnf.csiro.au/people/pulsar/psrcat} \citep{man05}). The spin frequency and LIGO upper limits are from \citet{abbott07}. The predicted maximum strain amplitudes are calculated using the distance and spin frequency of each of the pulsars, as well as the quadrupole as calculated using the results from the simulations. }
\end{center}
\end{table}

With the mechanical properties calculated in the molecular dynamics simulations it is found  that if PSR J2124-3358 was maximally deformed than it would have been detected as a source of gravitational waves. Assuming that the crustal strain is uniformly distributed to the breaking point, the non-detection of gravitational waves from this neutron star suggests that  either  the neutron star crust is not a perfect BCC crystal, or that the crust is not maximally deformed. The search for gravitational waves from pulsars could be used as a method to determining the structure, strength, or degree of deformation of the neutron star crust. 

\subsection{Bursts from Soft Gamma-ray Repeaters}\label{sec:bursts-from-soft}
Soft Gamma-ray Repeaters (SGRs) are characterized by their short bursts of low-energy gamma rays, which can reach a peak luminosity of $10^{41}\,\mathrm{erg\; s^{-1}}$ \citep{woods06}. There  have been eleven SGRs observed; seven confirmed and four candidates\footnote{Data from the McGill SGR/AXP online catalog, {\tt http://www.physics.mcgill.ca/{\kern -.15em\lower .7ex\hbox{\~{}}\kern .04em}pulsar/magnetar/main.html}}. Of these eleven SGRs three have been also observed to exhibit giant flare events, which can reach luminosities of $10^{44}\,\mathrm{erg/s}$ \citep{helfand79}. The observed SGR bursts have been associated with crustal activity; the bursts may be triggered by the yielding events due to stresses placed on the crust by the magnetic field \citep{chamel08}. The strength of the magnetic field required to trigger a burst is dependent on the breaking strain of the crust: 
\begin{equation}
\label{lfracture}
B = 10^{15}\left( \frac{E_{SGR}}{10^{41}\rm{erg}}\right)^{-1/2}\left(\frac{l}{\rm{km}}\right) \left(\frac{\phi_m}{10^{-3}}\right)\,\rm{G},
\end{equation} 
where $E_{SGR}$ is the energy of the burst, $l$ is the length of the crustal fracture and $\phi_m$ is the breaking strain \citep{thompson95}.  Using the results from the simulations, the observed energy emitted during burst and a neutron star's magnetic field strength, limits can be placed on the size of the crustal fracture. By estimating the fracture size, this can be used as an indication of if the burst can be attributed to the crust alone. If the fracture length required for an observed energy of a burst is larger than the crust size, then the burst can not be due to crustal events alone. 

The length of the fracture, using Equation \ref{lfracture}, can be expressed as: 
\begin{equation}
l = \left(\frac{B}{10^{15}\,\rm{G}}\right) \left(\frac{E_{SGR}}{10^{41}\,\rm{erg}}\right) \left(\frac{10^{-3}}{\phi_m}\right)\,\rm{km}.
\end{equation}
For a typical burst energy of $E_{SGR} = 10^{41}\,\rm{erg}$, assuming a magnetic field of $B = 10^{15}\,\rm{G}$, and using the simulation result of a yielding strain of 0.1 the corresponding fracture size is $l = 10\,\mathrm{m}$. In the case of the more energetic giant flares with $E_{SGR} = 10^{44}\,\rm{erg}$ the corresponding fracture length is $l = 300\,\rm{m}$. The neutron star crust has a depth of approximately $1\,\rm{km}$ with the outer crust encompassing ${\sim}100\,\rm{m}$. The fracture length associated with typical burst energies, $l = 10\,\rm{m}$, is within the confines of the crust, thus these events could be attributed to crustal activity. The giant flare events require a larger fracture length, $l = 300\,\rm{m}$, which would encompass ${\sim}15-30\%$  of the crustal depth. This larger fracture could possibly propagate along the surface or vertically and is within the size constraints of the neutron star crust. 

The crustal fracture lengths were calculated using properties of a perfect BCC lattice structure. The simulations with an imperfect crystal structure were found to display different mechanical properties from the perfect crystal structure simulations. A neutron star crust which yields at smaller strains would require a larger fracture length. Depending on the structure of the neutron star crust, how close it is to a perfect BCC lattice, crustal events could be ruled out as the mechanism for SGR bursts. 

\section{Conclusions}
\label{sec:conclusions}
In order to investigate the mechanical properties of a non-accreting neutron star crust we have performed various molecular dynamics simulations  using the software LAMMPS.  With these simulations tests were first performed in order to test the size effects of the simulation box size, as well as  determining the melting temperature of the various crustal compositions. A simple shear was applied to the crustal compositions by deforming the initially perfect BCC lattice simulation box of four different crustal compositions, those appropriate for a neutron star cooled via modified Urca, with a thick crust, and with a thin crust, as well as a pure iron crust. Additional simulations were performed to compare the stress-strain relationship of a system initially with a BCC  lattice structure to that which included defects. The temperature dependence on the the breaking strain as well as  the shear modulus was also investigated with additional simple shear simulations at different temperatures. The possibility of a second yielding event as well as the reversibility of the yielding event were also investigated in the four cases.  The calculation of the breaking strain and shear modulus from the simulations was applied to the context of gravitational wave emission from a neutron star due to a deformation on the neutron star surface, such a mountain. 

The mechanical properties for the different crustal compositions, which  include pure and impure compositions, with an initial BCC lattice structure were found to share similar characteristics, this includes the pure and impure simulations. The shear modulus at a density of $\rho = 10^{14}\,\mathrm{g\,cm^{-3}}$ was found to be around $1.78\times10^{30}\,\mathrm{dyne\,cm^{-2}}$ and the breaking strain was found to be ${\sim 0.1}$ in all four cases. This breaking strain measurement is in agreement with that found for an accreted neutron star crust composition \citep{horowitz09b}. As the shear was only applied in one crystallographic orientation, deforming along a different axis could result in different measured properties.  With these mechanical properties the constraints on the detection of gravitational waves were considered. The properties of three pulsars were considered in order to predict the strain amplitude associated with a neutron star with the calculated mechanical properties. LIGO had placed upper limits on the gravitational wave emission of 78 different pulsars and of the three investigated, one was found to have a strain amplitude larger then the upper limit placed on the neutron star. This indicates that if the crust had these calculated properties and was maximally deformed, then gravitational waves would have already been observed with LIGO from PSR J2124-3358. In the case of bursts from SGRs the fracture lengths required for observed burst energies were found to be within the confines of the neutron star crust. 

Crustal material which included defects as compared to that which had an initially perfect BCC lattice structure were found to have differing stress-strain relationships. The defects in the simulations were introduced by first melting the perfect BCC lattice and then cooling the material to a solid. This behaviour of the initially imperfect crystal structure was found to display a similar shear modulus to that of the perfect BCC lattice crystal after the first major yielding event occurs. The effect of defects on an accreted neutron star crust were examined in \citet{horowitz09b} by introducing six grains to the simulation. The addition of defects in the accreted crust case were found to affect  the stress-strain relationship only moderately. The differences between the defects in the non-accreted and accreted crust composition could be due to the number of grains in each of the simulations. Future work would include investigating the affect of grain number on the corresponding stress-strain relationship. It should be noted that the crustal composition in the accreted case has an impurity factor which is higher than the limits placed on the factor through neutron star cooling \citep{horowitz09b}. A neutron star which has not undergone any accretion events is unlikely, and in the accreted crust case the impurity factor is higher than observational constrains, the results from these two cases may represent two extremes of what may be occurring in nature. In a low mass X-ray binary systems with an accretion rate of $10^{-9}\mathrm{\msun\,yr^{-1}}$ the crust could be replaced via accretion in $10^7$ years \citep{chamel08}. In the case of dim isolated neutron stars, as well as AXPs and SGRs the accretion rate has been estimated to range from $3.2\times10^{14}\mathrm{g\,s^{-1}}$ to $4.2\times10^{17}\mathrm{g\,s^{-1}}$ or in solar mass units the accretion rates are $5\times10^{-12}\mathrm{\msun\,yr^{-1}}$ to $7\times10^{-9}\mathrm{\msun\, yr^{-1}}$ \citep{alpar01}. For an isolated neutron star these accretion rates correspond to a total crustal mass of $0.01\msun$ \citep{chamel08} being replaced in ${\sim}10^{9}$ to ${\sim}10^{6}$years. The accretion rate can also be estimated by attributing the observed luminosity to accretion. For at 10\,km and 1.4\msun dim isolated neutron star, with a luminosity of $10^{32}\rm{erg}$, this corresponds to $\dot{M} \sim 10^{-14}\msun\,yr^{-1} $ and the crust being replaced by accretion in $10^{12}$ years. As such, the crustal composition calculated for  the non-accreting neutron stare, presented in this paper, would hold for those isolated neutron stars which have an accretion rate on the lower end of the rates.

The molecular dynamics simulations reported here leave many venues for future work  in order to fully understand the neutron star crust. The neutron star crust would not be expected to have a perfect lattice structure, thus it would be important to understand the effect of defects added to the simulation, as well how the orientation of the applied shear also affects the material. Fully characterizing the material as the simulation progresses would give an indication of how the structure of the crustal material changes with applied shear and cooling, this may have an effect on the material properties, as well as conductivity, of the crust. In all the simulations magnetic fields were not directly placed within the simulation, the addition of magnetic fields may have an effect on the mechanical properties, especially as with high magnetic fields there is a change in the electron screening \citep{rishi11}. The simulation results reported here indicate upper limits on the crustal properties, further simulations may bring us closer to understanding the true nature of the neutron star crust. 

\section*{Acknowledgments}

This research was supported by funding from NSERC. The calculations
were performed on computing infrastructure purchased with funds from
the Canadian Foundation for Innovation and the British Columbia
Knowledge Development Fund.

\bibliographystyle{mn2e}
\bibliography{complete}

\begin{thebibliography}{}

\bibitem[\protect\citeauthoryear{{Abbott}, {Abbott}, {Adhikari}, {Agresti},
  {Ajith}, {Allen}, {Amin}, {Anderson}, {Anderson}, {Arain} \& et al.}{{Abbott}
  et~al.}{2007}]{abbott07}
{Abbott} B.,  {Abbott} R.,  {Adhikari} R.,  {Agresti} J.,  {Ajith} P.,  {Allen}
  B.,  {Amin} R.,  {Anderson} S.~B.,  {Anderson} W.~G.,  {Arain} M.,    et al.
  2007, \prd, 76, 042001

\bibitem[\protect\citeauthoryear{{Alpar}}{{Alpar}}{2001}]{alpar01}
{Alpar} M.~A.,  2001, \apj, 554, 1245

\bibitem[\protect\citeauthoryear{{Chamel} \& {Haensel}}{{Chamel} \&
  {Haensel}}{2008}]{chamel08}
{Chamel} N.,  {Haensel} P.,  2008, Living Reviews in Relativity, 11, 10

\bibitem[\protect\citeauthoryear{{Gupta}, {Brown}, {Schatz}, {M{\"o}ller} \&
  {Kratz}}{{Gupta} et~al.}{2007}]{gupta07}
{Gupta} S.,  {Brown} E.~F.,  {Schatz} H.,  {M{\"o}ller} P.,    {Kratz} K.,
  2007, \apj, 662, 1188

\bibitem[\protect\citeauthoryear{{Helfand} \& {Long}}{{Helfand} \&
  {Long}}{1979}]{helfand79}
{Helfand} D.~J.,  {Long} K.~S.,  1979, Nature, 282, 589

\bibitem[\protect\citeauthoryear{{Hoffman} \& {Heyl}}{{Hoffman} \&
  {Heyl}}{2009}]{hoffman09}
{Hoffman} K.,  {Heyl} J.,  2009, \mnras, 400, 1986

\bibitem[\protect\citeauthoryear{{Horowitz} \& {Kadau}}{{Horowitz} \&
  {Kadau}}{2009}]{horowitz09b}
{Horowitz} C.~J.,  {Kadau} K.,  2009, Physical Review Letters, 102, 191102

\bibitem[\protect\citeauthoryear{{Itoh} \& {Kohyama}}{{Itoh} \&
  {Kohyama}}{1993}]{itoh93}
{Itoh} N.,  {Kohyama} Y.,  1993, \apj, 404, 268

\bibitem[\protect\citeauthoryear{{Lattimer} \& {Prakash}}{{Lattimer} \&
  {Prakash}}{2004}]{lattimer04}
{Lattimer} J.~M.,  {Prakash} M.,  2004, Science, 304, 536

\bibitem[\protect\citeauthoryear{{Lattimer}, {van Riper}, {Prakash} \&
  {Prakash}}{{Lattimer} et~al.}{1994}]{latt94}
{Lattimer} J.~M.,  {van Riper} K.~A.,  {Prakash} M.,    {Prakash} M.,  1994,
  \apj, 425, 802

\bibitem[\protect\citeauthoryear{{Manchester}, {Hobbs}, {Teoh} \&
  {Hobbs}}{{Manchester} et~al.}{2005}]{man05}
{Manchester} R.~N.,  {Hobbs} G.~B.,  {Teoh} A.,    {Hobbs} M.,  2005, \aj, 129,
  1993

\bibitem[\protect\citeauthoryear{{Nadal} \& {Le Poac}}{{Nadal} \& {Le
  Poac}}{2003}]{npref}
{Nadal} M.-H.,  {Le Poac} P.,  2003, Journal of Applied Physics, 93, 2472

\bibitem[\protect\citeauthoryear{{Plimpton}}{{Plimpton}}{1995}]{plimpton95}
{Plimpton} S.,  1995, Journal of Computational Physics, 117, 1

\bibitem[\protect\citeauthoryear{{Ruderman}}{{Ruderman}}{1969}]{ruderman69}
{Ruderman} M.,  1969, Nature, 223, 597

\bibitem[\protect\citeauthoryear{{Schneider} \& {Stoll}}{{Schneider} \&
  {Stoll}}{1978}]{langevin}
{Schneider} T.,  {Stoll} E.,  1978, \prb, 17, 1302

\bibitem[\protect\citeauthoryear{{Sharma} \& {Reddy}}{{Sharma} \&
  {Reddy}}{2011}]{rishi11}
{Sharma} R.,  {Reddy} S.,  2011, \prc, 83, 025803

\bibitem[\protect\citeauthoryear{{Smoluchowski}}{{Smoluchowski}}{1970}]{smoluchowski70}
{Smoluchowski} R.,  1970, Physical Review Letters, 24, 923

\bibitem[\protect\citeauthoryear{{Swope}, {Andersen}, {Berens} \&
  {Wilson}}{{Swope} et~al.}{1982}]{vverlet}
{Swope} W.~C.,  {Andersen} H.~C.,  {Berens} P.~H.,    {Wilson} K.~R.,  1982,
  \jcp, 76, 637

\bibitem[\protect\citeauthoryear{{Thompson} \& {Duncan}}{{Thompson} \&
  {Duncan}}{1995}]{thompson95}
{Thompson} C.,  {Duncan} R.~C.,  1995, \mnras, 275, 255

\bibitem[\protect\citeauthoryear{{Timmes}, {Hoffman} \& {Woosley}}{{Timmes}
  et~al.}{2000}]{timmesapjs00}
{Timmes} F.~X.,  {Hoffman} R.~D.,    {Woosley} S.~E.,  2000, \apjs, 129, 377

\bibitem[\protect\citeauthoryear{{Ushomirsky}, {Cutler} \&
  {Bildsten}}{{Ushomirsky} et~al.}{2000}]{gregu}
{Ushomirsky} G.,  {Cutler} C.,    {Bildsten} L.,  2000, \mnras, 319, 902

\bibitem[\protect\citeauthoryear{{Verlet}}{{Verlet}}{1967}]{verletref}
{Verlet} L.,  1967, Physical Review, 159, 98

\bibitem[\protect\citeauthoryear{{Woods} \& {Thompson}}{{Woods} \&
  {Thompson}}{2006}]{woods06}
{Woods} P.~M.,  {Thompson} C.,  2006, in {Lewin, W.~H.~G.~\& van der Klis, M.}
  ed., Compact stellar X-ray sources {Soft gamma repeaters and anomalous X-ray
  pulsars: magnetar candidates}.
{Cambridge: Cambridge Univ. Press.}, pp 547--586

\end{thebibliography}

\label{lastpage}

\end{document}